\documentclass[12pt,preprint]{aastex}
\newcommand{\be}{\begin{equation}}
\newcommand{\ee}{\end{equation}}
\newcommand{\kms}{\,\hbox{km s}^{-1}}
\newcommand{\km}{\,\hbox{km}}
\newcommand{\cm}{\,\hbox{cm}}
\newcommand{\yr}{\,\hbox{y}}
\newcommand{\half}{{\textstyle{1\over2}}}

\def\ffrac#1#2{{\textstyle\frac{#1}{#2}}}
\def\bfx{{\bf x}}
\def\bfv{{\bf v}}

\def\p{\partial}
\def\bfe{{\bf e}}

\shorttitle{Irregular structure in Saturn's rings}
\shortauthors{Tremaine}

\begin{document}

\title{On the origin of irregular structure in Saturn's rings} 

\author{Scott Tremaine}

\affil{Princeton University Observatory, Peyton Hall,
Princeton, NJ 08544}
\email{tremaine@astro.princeton.edu}

\begin{abstract}

We suggest that the irregular structure in Saturn's B ring arises from the
formation of shear-free ring-particle assemblies of up to $\sim 100\km$ in
radial extent. The characteristic scale of the irregular structure is set by
the competition between tidal forces and the yield stress of these assemblies;
the required tensile strength of $\sim 10^5\hbox{ dyn cm}^{-2}$ is consistent
with the sticking forces observed in laboratory simulations of frosted ice
particles. These assemblies could be the nonlinear outcome of a linear
instability that occurs in a rotating fluid disk in which the shear stress is
a decreasing function of the shear. We show that a simple model of an
incompressible, non-Newtonian fluid in shear flow leads to the Cahn-Hilliard
equation, which is widely used to model the formation of structure in binary
alloys and other systems.

\end{abstract}


\section{Introduction}

The Voyager 1 and 2 spacecraft, which flew past Saturn in 1980 and 1981,
revolutionized our understanding of the Saturn system. One of the
remarkable features discovered by Voyager was rich radial structure in
Saturn's rings. Although some of this structure, mostly in the outer or A
ring, is known to arise from density or bending waves generated by discrete
resonances with the inner satellites, most of the radial structure remains
unexplained, particularly in the main or B ring. \citet{hc96} point out that
wavetrains associated with known resonances cover less than 1\% of the radial
extent of the A and B rings. The remaining vast majority of the structure is
often called ``irregular'' since the bright and dark features show little or
no long-range coherence.

The purpose of this paper is to suggest and examine a novel explanation for
irregular structure in Saturn's rings. Ring particles are likely to have
(weak) cohesive forces, and therefore can assemble or ``freeze'' into
structures much larger than an individual ring particle. The size of these
structures is limited by tidal forces from Saturn and collisional erosion by
impacting particles. We suggest that much of the irregular structure may
consist of alternating annuli of ``solid'' and ``liquid'' ring material, the
former consisting of an assembly of ring particles frozen into rigid rotation
around Saturn, and the latter consisting of individual ring particles in
differential rotation. Thus we hypothesize that the irregular structure is
primarily a manifestation of variations in shear rather than surface density.

Rich spatial structure associated with the co-existence of two phases is a
feature of many physical processes, including ferromagnetism, spinodal
decomposition in alloys, crystal growth, chemical reactions, and even traffic
flow. We shall exploit some of these analogies in \S\ref{sec:toy}. One
complication in Saturn's rings is that the physics is non-local, because the
tidal stresses on a rigidly rotating annulus depend on its total radial
extent. We estimate these stresses and the corresponding upper limit to the
radial extent of rigidly rotating annuli in \S\ref{sec:tensile}.

We begin by reviewing the observations of the irregular structure in
\S\ref{sec:obs} and competing theoretical explanations in \S\ref{sec:theory}.
In \S\ref{sec:four} we review the equations that describe the dynamics of the
ring fluid. In \S\ref{sec:toy} we describe a simple toy model that illustrates
some of the important dynamical behavior associated with the irregular
structure, and sets the model in the context of the theory of phase
transitions.

The suggestion that some areas in planetary rings are locked into solid
assemblies is not new to this paper. \citet{wt88} point out that dense rings
may have a liquid-solid phase transition; they did not find any such
transition in their numerical experiments on inelastically colliding hard
spheres, but this is not surprising since they did not model any cohesive
forces. \citet{mos02} suggest that ring-particle shocks or ``jams'' may occur
in sectors of converging flow in an eccentric ring; however, our focus is on
axisymmetric ring-particle structures held together by cohesive forces, while
theirs is on non-axisymmetric structures held together by self-gravity and ram
pressure.

\section{Observations}

\label{sec:obs}

Voyager images reveal irregular structure throughout the A and B rings, over a
wide range of scales ranging from several hundred km to the resolution limit
of about 5 km. In regions where the ring is sufficiently transparent, the
surface-brightness variations track the optical-depth variations detected
during stellar occultations observed by Voyager
\citep{hc96}. However, in the main part of the B ring the optical depth is so
high that there is little or no signal from the occultation experiments;
moreover, the brightness variations in this region cannot be explained by
optical depth variations, since the solar elevation at the time of the
encounter was so low ($\lesssim 8^\circ$) that variations in optical depth
have negligible influence on the surface brightness in reflected light. Thus,
other effects such as particle albedo or phase function, which may or may not
be correlated with optical depth variations, must be responsible for most of
the irregular structure in the B ring. Comparison of images at different phase
angles \citep{cuz84} suggests that some features are due to variations in
albedo and others due to variations in phase function. The surface-brightness
ratios at different phase angles appear to be bimodal, suggesting that there
are only two distinct phase functions.

Most of the irregular structure appears to be axisymmetric, at least on scales
$\gtrsim 50\km$ and away from major resonances. At smaller scales the features
at the same radius and different longitudes cannot be matched, implying that
they are non-axisymmetric or time-variable or both.

\citet{hc96} have measured the local power spectrum of the reflectivity of the
B ring as a function of radius (from 92,000 km at its inner edge to 122,000 km
at the Cassini division). In most regions the dominant wavelength varies
between 100 and 200 km. In the outer 1500 km of the ring, and a few other
isolated regions, there is substantial power at wavelengths as short as tens
of km. There are also regions of 1000 km or so in which little or no irregular
structure is visible.

The particle-size distribution in Saturn's rings is constrained by the Voyager
occultation experiments \citep{zeb85,sho90}. These observations suggest a broad
distribution of particle sizes with an upper cutoff of a few meters, but
cannot distinguish whether these particles are in differential rotation or
locked by contact forces into a solid assembly. 

\section{Theoretical models}

\label{sec:theory}

The irregular structure is surprising because viscous diffusion is expected to
smooth out such structure on timescales much less than the age of the
rings. The characteristic time required to smooth out structure on a radial
scale $\Delta r$ is
\be
t_\nu={(\Delta r)^2\over\nu}=3\times 10^4\yr\left(\Delta r\over
100\km\right)^2\left(100\cm^2\sec^{-1}\over\nu\right),
\ee
where $\nu$ is the kinematic viscosity, which can be estimated either from the
rate of damping of density waves, or from kinetic theory together with
estimates of the particle size and optical depth. The very small value of
$t_\nu$ implies that all small-scale radial structure should be erased
from the B ring, unless it is actively maintained by some mechanism.

One possible explanation is that the irregular structure arises from
gravitational shepherding by small moonlets or large particles. However,
objects large enough to influence the ring over scales $\gtrsim 100\km$
should also clear a gap around themselves, and such gaps are not seen
\citep{eow83}. 

One attractive hypothesis is that the irregular structure arises from local
axisymmetric instabilities in the rings. There are two natural candidates for
this instability:

\begin{itemize}

\item Viscous instability. In most models for axisymmetric ring dynamics, the
kinematic viscosity $\nu$ is assumed to be a function only of the surface
density $\mu$. The viscous instability then arises if the angular-momentum
flux in the rings is a decreasing function of surface density, that is, if
\be
K_1\equiv -{d(\mu\nu)\over d\mu} > 0;
\label{eq:visc}
\ee
the growth rate of the instability is $\gamma=3k^2K_1$, where $k$ is the
radial wavenumber. This is a secular instability, since the growth rate is
proportional to the strength of the dissipative forces (i.e.,
$\gamma\propto\nu$). If condition (\ref{eq:visc}) is satisfied, ring material
preferentially migrates from regions of low surface density to regions of high
surface density, thereby adding mass to the high-density regions, and
depleting the low-density regions even further.

The viscous instability was first discussed in the context of accretion disks
by \citet{le74}, and was invoked to explain the irregular structure in
Saturn's rings by \citet{lb81}, \citet{luk81}, and \citet{war81}. These
authors were motivated by the observation that the viscosity in dilute rings
appears to satisfy the instability condition (\ref{eq:visc}). However, the B
ring is unlikely to be dilute, unless the elasticity of its constituent
particles is unrealistically high. Both analytic kinetic theory and $N$-body
simulations of the viscosity in a dense ring composed of inelastic hard
spheres show that the instability condition (\ref{eq:visc}) is not satisfied
by a wide margin \citep{at86,wt88}; therefore, the viscous instability is
unlikely to operate in Saturn's B ring.

\item Viscous overstability. Density waves can propagate in planetary rings,
and dozens of density wavetrains have been identified in Saturn's A ring. The
collective effect that dominates density-wave propagation is the self-gravity
of the ring. Self-gravity is important, even though the ring is much less
massive than the planet, because the thickness of the ring is also much less
than its radius. A more precise statement is that Toomre's $Q$ parameter is of
order unity in the rings. Viscosity can lead to either decay or growth of
density waves; the latter case, known as viscous overstability, occurs if
\be
K_2\equiv 3\mu{d\nu\over d\mu}+\ffrac{2}{3}\nu-\xi > 0;
\label{eq:visco}
\ee
here $\nu$ and $\xi$ are the kinematic shear and bulk viscosity, and both are
assumed to depend only on the surface density $\mu$. The growth rate is
$\gamma= \half k^2K_2$. Unfortunately, the assumptions that the stress tensor
can be represented by a shear and bulk viscosity, and that these depend only
on the surface density, are harder to justify than in the case of the viscous
instability, since an overstable ring oscillates on a timescale comparable to
the interparticle collision time.

The existence of this instability was first pointed out by \citet{kat78} for
accretion disks (where the important collective effect is pressure rather than
self-gravity), and by \citet{gt78} for planetary rings. \citet{bgt85} describe
a crude kinetic theory for a ring composed of inelastic, closely packed, hard
spheres, and show that the viscous instability is present in this
approximation. \citet{st95} discuss in detail the local linear stability of a
differentially rotating, two-dimensional, isothermal, fluid disk, in which the
viscosity depends only on surface density, and the effects of self-gravity are
included; they derive a cubic dispersion relation that contains the viscous
instability and viscous overstability as special cases. \citet{st99} describe
numerical calculations of the nonlinear evolution of these disks, which show
that the viscous overstability can lead to a rich radial structure with
surface-density contrast of order unity. Numerical experiments on
self-gravitating collections of ring particles appear to exhibit a viscous
overstability in some cases \citep{mos96,dai01,sss01}. However, the most
unstable wavelengths in the linear calculations, and the wavelengths with
significant power in the nonlinear calculations, are at most a few hundred
meters---almost three orders of magnitude smaller than the dominant
wavelengths in the irregular structure in the B ring
\citep{hc96}. \citet{st99} argue that nonlinear wave-wave interactions
transfer the fluctuation power to much larger wavelengths, but so far this
suggestion remains untested.

\end{itemize}

A quite different proposal is that some of the irregular structure arises from
unstable ballistic transport \citep{lis84,dur89,dur92,dur95}. Impacts by
interplanetary particles erode the ring particles and redistribute their mass
to adjacent annuli. This process is unstable because high-density regions of
the ring tend to absorb more of the ejecta than neighboring regions. However,
the growth rate of the instability is negligible for optical depths $\gtrsim
1.5$. Thus, ballistic transport is probably unable to explain the irregular
structure in the bulk of the B ring.

\section{Shear stress in dense planetary rings}

\label{sec:four}

In this section we investigate the dynamics, stresses, and stability of solid
and liquid ring phases.

\subsection{Equations of motion}

\label{sec:eqmot}

We examine a ring orbiting a point mass $M$. Test particles on circular orbits
travel at the Keplerian angular speed $\Omega_K(r)=(GM/ r^3)^{1/2}$. We employ
a rotating Cartesian coordinate system in which $\hat\bfe_x$ points radially
outward, $\hat\bfe_y$ points in the direction of rotation, and the origin, at
radius $R$, rotates around the central mass at angular speed
$\Omega\equiv\Omega_K(R)$. We work at distances from the origin that are small
compared to the orbital radius $R$ (Hill's approximation). We ignore the
vertical structure of the disk, treating it as a razor-thin sheet with surface
density $\mu(\bfx,t)$. We restrict ourselves to axisymmetric disturbances, so
that $\p/\p y=0$. The Euler and continuity equations then read
\begin{eqnarray}
{\p u\over \p t} + u{\p u\over\p x} &=& 3\Omega^2 x + 2\Omega v +{1\over
\mu}{\p \Sigma_{xx}\over \p x}, \nonumber \\ {\p v\over \p t} + u{\p v\over\p
x} &=&- 2\Omega u + {1\over\mu}{\p
\Sigma_{xy}\over \p x}, \nonumber \\
{\p\mu\over\p t} +{\p\over\p x}(\mu u) &=& 0,
\label{eq:motion}
\end{eqnarray}
where $\bfv(\bfx,t)=u(x,t)\hat\bfe_x+v(x,t)\hat\bfe_y$ is the velocity in the
rotating frame, and $\Sigma_{ik}=\int \sigma_{ik}dz$ is the vertically
integrated stress tensor (dimensions of force per unit length).  We define the
tangential shear to be
\be
s\equiv {\p v\over \p x}.
\label{eq:sheardef}
\ee

In a disk with zero stress gradients, a solution of the equations of motion
is
\be 
u(x,t)=0, \qquad v(x,t)=v_K(x)=-\ffrac{3}{2}\Omega x,
\label{eq:kepsol}
\ee
corresponding to circular Keplerian orbits and constant shear $s=s_K\equiv
-{3\over 2}\Omega$. In a solid disk with zero shear, a solution of the
equations of motion is
\be
u(x,t)=0, \quad v(x,t)=v_s=\hbox{constant}, \quad
\Sigma_{xy}=\hbox{constant}, \quad {\p\Sigma_{xx}\over\p x}=-\mu(2\Omega
v_s+3\Omega^2x).
\ee
If we assume that the surface density is constant, that the zero-shear region
extends from $x_1$ to $x_2$, and that the tensile stress vanishes at
the edges of the solid region (i.e. $\Sigma_{xy}=0$ at $x_1,x_2$) then the
last of these equations can be integrated to yield
\be
\Sigma_{xx}=\ffrac{3}{2}\mu\Omega^2(x-x_1)(x_2-x), \qquad
v_s=-\ffrac{3}{4}\Omega(x_1+x_2). 
\ee
Thus, any solid annulus is subject to a tensile stress, the maximum of which
occurs at its midline and is equal to 
\be
\Sigma_{xx,\rm max}=\ffrac{3}{8}\mu\Omega^2(\Delta x)^2,
\label{eq:sigmax}
\ee
where $\Delta x\equiv x_2-x_1$ is the width of the annulus. 

If we linearize the first two of equations (\ref{eq:motion}) with respect to a
state of uniform shear ($u_0=0$, $\p v_0/\p x=s$), and neglect perturbations in
the stress tensor, then small disturbances are stable if and only if
$s>-2\Omega$. This is the well-known Rayleigh criterion for the stability of
Couette flow \citep{cha61}.

\subsection{Tensile stress}

\label{sec:tensile}

Equation (\ref{eq:sigmax}) for the maximum height-integrated tensile stress
(force per unit length) in a solid annulus can be converted to an equation for
the ordinary tensile stress (force per unit area) by approximating the
vertical structure of the ring as that of a homogeneous slab with thickness
$h$ and density $\rho=\mu/h$ (note that $\rho$ is less than the density of the
ring particles, by the filling factor):
\be
\sigma_{xx,\rm max}={\Sigma_{xx,\rm max}\over h}=
\ffrac{3}{8}\rho\Omega^2(\Delta r)^2=4\times 10^5\hbox{ dyn
cm}^{-2}\left(\rho\over0.3\hbox{ g cm}^{-3}\right)\left(10^{10}\cm\over
r\right)^3\left(\Delta r\over 100\km\right)^2,
\ee
in which we have inserted parameters appropriate for Saturn. The maximum width
of a solid annulus with a given tensile strength may be called its ``tidal
width''. If we identify the tidal width with half of the dominant wavelength
of 100 km seen in the irregular structure (i.e. $\Delta r=50\km$) then we
require that a frozen assembly of ring particles have a tensile strength or
yield stress $\gtrsim 1\times 10^5\hbox{ dyn cm}^{-2}$.

The strength of ring-particle assemblies is, of course, very difficult to
estimate. An upper limit is the yield stress of solid ice, which is
$\sigma_{\rm max}\sim 10^7\hbox{ dyn cm}^{-2}$ at temperatures 5--10 K below
freezing \citep{sch99}, and probably higher at the much lower temperatures
characteristic of Saturn's rings. However, the actual strength of ice-particle
assemblies can be many orders of magnitude lower: comets, which may be rubble
piles of icy particles, have yield stresses $\sigma_{\rm max}\lesssim 10^3
\hbox{ dyn cm}^{-2}$, and  \citet{ab96} derive an even lower strength
$\lesssim 10^2\hbox{ dyn cm}^{-2}$ for Comet Shoemaker-Levy 9.

Experimental work on sticking of cm-sized ice particles is reviewed by
\citet{bsl96}. The experiments show no sticking unless the ice surfaces are
coated with uncompacted frost; however, thin water-frost layers can lead to
sticking forces up to $10^4\hbox{ dyn}$ over contact areas $\sim 0.01\cm^2$
(see also Supulver et al.~1997). On Saturn's ring particles, frost is likely
to be produced by dust-particle impacts, and removed by collisions with other
particles; \citet{hbl91} conclude that centimeter-sized or larger particles
will be frost-coated, although this conclusion relies on a very uncertain
estimate of the optical depth in grains. Whether or not frost layers are
present on B-ring particles is therefore an open question; we shall assume that
frost is present in order to have significant sticking forces. 

If the sticking force scales with the contact area, then the forces found by
\citet{bsl96} would imply a tensile strength per unit area of $10^6\hbox{ dyn
cm}^{-2}$. This is certainly an overestimate, since the contact area
between ring particles is only a fraction of the total area.  If there is a
wide range of particle sizes, the contact points and the yield stress are
likely to be dominated by small particles, which provide a kind of ``cement''
to hold the large particles together. We now make a crude estimate of the
yield stress. For spherical particles of radius $R$, the contact area is
$A=2\pi R d$ where $d$ is the thickness of the frost layer. We assume that the
contact force is proportional to the contact area, that the force is $\approx
10^4\hbox{ dyn}$ for a contact area $A=0.01\cm^2$, that the thickness of the
frost layer is $d\approx10^{-2}\cm$, and that there are $N\approx 1\cm^2/\pi
R^2$ contacts per cm$^2$. Then the yield stress is $\sigma_{\rm max}\approx
2\times 10^4 (1\cm/R)\hbox{ dyn cm}^{-2}$, comparable to the required yield
stress if $R\approx 0.2\cm$. Note that most of the mass of the ring could
be---and likely is---in much larger particles, while most of the sticking
force comes from the small particles. 

Therefore, within the very large uncertainties, it is plausible that annular
ring-particle assemblies as large as $\Delta r\sim 50$--100 km could have the
required strength ($\sigma_{\rm max}\sim10^5\hbox{ dyn cm}^{-2}$) to survive
tidal stresses.

We now briefly discuss the orbital stability of solid annuli. Laplace and
later Maxwell showed that a solid annulus orbiting a planet is unstable, with
growth time $\Omega/\surd{2}$. However, \citet{fmp84} point out that the
Laplace-Maxwell analysis assumes that the annulus is completely rigid, in
particular that the internal sound speed $c\gg\Omega r$. In fact $c\ll\Omega
r\simeq 20\kms$ for any plausible material. In this case the nature of the
instability changes dramatically \citep{fmp84}: the growth rate becomes much
smaller, $\sqrt{3}mc/r$, where $m$ is the azimuthal wavenumber; the
instability is predominantly azimuthal; and the $m=1$ instability in the rigid
ring is replaced by an instability that is present for all $m$ and grows
faster as $m$ increases. The physical basis for the instability is simple: if
a mass element moves closer to the element in front of it, it is pushed
backwards by elastic forces. Thus it loses angular momentum, its angular speed
increases, and it moves even closer to the element in front.

One limitation of the analysis of \citet{fmp84} is the neglect of the rigidity
of the ring: they assume that the modulus of compression is non-zero but the
modulus of rigidity $\mu$ is zero. Rigidity suppresses the instability at
short wavelengths, for $m>m_{\rm crit}\equiv (3\rho/\mu)^{1/2}\Omega r$, where
$\rho$ is the density in the ring. 

Thus there are several possibilities that may enable solid ring-particle
assemblies to be stable. Possibly the rigidity is large enough to stabilize
all azimuthal wavenumbers; possibly the solid regions are arcs, rather than
annuli, spanning angles $\lesssim 2\pi/m_{\rm crit}$; possibly there is a
slowly growing instability that disrupts the solid annuli, so that they are
constantly dissolving and re-forming; or perhaps other effects such as
pressure from adjacent parts of the ring can stabilize a solid annulus.
We cannot say which of these possibilities is more likely, but it would be
premature to dismiss the possibility of solid annuli in the ring on the basis
of stability arguments. 

\subsection{Shear stress}

The simplest model for the shear stress is $\Sigma_{xy}=E s$, where
$E=\int \eta dz$, the vertically integrated viscosity, is assumed to be a
function only of the surface density $\mu$. Many authors have investigated the
stability and evolution of models of this kind
\citep{le74,gt78,kat78,lb81,st95,st99}. However, granular systems generally
behave as non-Newtonian fluids, in which the viscosity is a function of the
shear, so that the shear stress $\Sigma_{xy}$ is likely to be a nonlinear
function of the shear.

For example, a simple kinetic theory for dense planetary rings is described by
\citet{bgt85}. In their model, the shear stress and radial stress are
\be
\Sigma_{xy}=q_1F{\mu^3\Omega^2\over \rho^2}\hbox{sgn}\,(s),\qquad
\Sigma_{xx}=-q_2F{\mu^3\Omega^2\over\rho^2};
\label{eq:bgt}
\ee
here $q_1$ and $q_2$ are dimensionless constants of order unity, and $\rho$ is
the density of the ring particles or the mean density inside the ring (the
theory is not accurate enough to distinguish these two densities, as it
assumes in effect that the ring is incompressible). The dimensionless factor
\be
F=1+{4\pi G\rho\over\Omega^2}=1+6.7\left(\rho\over 0.3
\hbox{ g cm}^{-3}\right)\left(r\over 10^{10}\cm\right)^3
\label{eq:fdef}
\ee
is the enhancement in the vertical gravitational field due to
the self-gravity of the ring. The negative radial stress $\Sigma_{xx}$
represents the hydrostatic pressure required to maintain the ring thickness. 

More generally, the shear stress is likely to be a complex function of the
shear, the surface density, and perhaps also their histories. Analytic models
for the constitutive relations of a dense ring of inelastically colliding
particles are still in a primitive state, particularly when cohesive forces
are present. On the other hand $N$-body simulations of ring dynamics, which
will be an indispensable guide to the correct analytic theory, have so far
been restricted to the Keplerian shear rate and do not include the possibility
of cohesion.

Given our present ignorance, the only practical approach is to parametrize a
fairly general set of constitutive relations, and investigate the behavior of
the ring as a function of these parameters. We shall therefore assume that the
shear stress and the radial stress are arbitrary nonlinear functions of the
shear $s$ and the surface density $\mu$. We can deduce some plausible general
properties of these stresses. The shear stress $\Sigma_{xy}(\mu,s)$ should
vanish if the shear vanishes (if there is no shear then there is no
stress). We also expect that $\Sigma_{xy}$ will become very large as the shear
$s$ approaches $-2\Omega$, since we have seen in \S\ref{sec:eqmot} that
circular orbits are unstable when $s<-2\Omega$. It is also natural to assume
that $\Sigma_{xy}$ has the same sign as the shear $s$; however, we do not
assume that $\Sigma_{xy}$ is monotonic in $s$, and in subsequent sections will
focus on instabilities that can arise when $\Sigma_{xy}$ is a decreasing
function of $s$ in some interval. The radial stress $\Sigma_{xx}(\mu,s)$ will
be negative if the ring behaves as a fluid that exerts a pressure force, but
may become positive in regions of negligible shear when the ring material
freezes.

\subsection{Linear stability}

\label{sec:linstab}

As described above, we shall assume that the stresses are functions of the
local shear and surface density,
\be
\Sigma_{xx}=f(\mu,s)\qquad,\qquad \Sigma_{xy}=g(\mu,s).
\ee
We then linearize the equations of motion (\ref{eq:motion}) around a state in
which the surface density is uniform, the unperturbed motion is Keplerian
(eq.\ \ref{eq:kepsol}), and the stresses are constant. We assume that the
perturbations vary as $\exp(ikx+\gamma t)$.  We find
\begin{eqnarray}
\gamma u_1 -2\Omega v_1 & = & {ik\over\mu_0}\left(f_\mu \mu_1 +
ikf_s v_1\right), \nonumber \\
\gamma v_1 + \half\Omega u_1 & = & {ik\over\mu_0}\left(g_\mu \mu_1
+ikg_s v_1\right),
\nonumber \\
\gamma \mu_1 +ik\mu_0u_1 & = & 0,
\end{eqnarray}
where $f_s\equiv(\p f/\p s)_0$, etc. The resulting dispersion relation
is\footnote{This analysis neglects the effects of the ring self-gravity; these
can easily be added to (\ref{eq:disprel}) and have no important effect on our
conclusions.}
\be
\mu\gamma^3 +k^2g_s\gamma^2+(\Omega^2\mu -k^2\mu f_\mu -\half\Omega
k^2f_s)\gamma + k^4(f_sg_\mu-f_\mu g_s) - 2\Omega k^2 \mu g_\mu = 0.
\label{eq:disprel}
\ee

We can recover the viscous instability and overstability of \S\ref{sec:theory}
by neglecting the radial stress ($f=0$) and assuming that the ring is a
Newtonian fluid so that $g=\mu s \nu(\mu)$. Equation (\ref{eq:disprel}) then
simplifies to
\be
\gamma^3+k^2\nu\gamma^2+\Omega^2\gamma+3\Omega^2k^2{d(\mu\nu)\over
d\mu}=0.
\ee
If the viscosity is small, this equation has two possible solutions: either
$\gamma=-3k^2d(\mu\nu)/d\mu+\hbox{O}(\nu^2)$, corresponding to the viscous
instability (eq.\ \ref{eq:visc}), or $\gamma=\pm i\Omega+k^2[{3\over 2}\mu
(d\nu/d\mu)+\nu]$, corresponding to the viscous overstability (eq.\
\ref{eq:visco}, with the bulk viscosity $\xi=-{4\over 3}\nu$ so that the
radial stress vanishes).

Thus the viscous instability and overstability arise in the limit where the
stresses $f,g\to 0$ at fixed wavenumber $k$. In this paper, we will focus
instead on the short-wavelength limit, in which $k\to\infty$ at fixed values
of $f,g$. In this case the roots of the dispersion relation are given by
\be
\gamma=-{k^2g_s\over\mu}+\hbox{O}(k^0),\qquad
\gamma^2=k^2\left(f_\mu-f_s{g_\mu\over g_s}\right) + \hbox{O}(k^0).
\label{eq:groot}
\ee
The first of these is the largest in absolute value, and leads to a rapidly
growing instability if and only if $g_s<0$. This instability does not arise in
a Newtonian disk, in which $g=E s$ so that $g_s=E>0$.

Assuming that the stresses $f$ and $g$ are comparable in magnitude, and that
$f_s\sim f/\Omega$ and $f_\mu\sim f/\mu$, with similar relations for $g$, the
approximations that lead to the formula (\ref{eq:groot}) are valid so long as
$k^2g\gtrsim \Omega^2\mu$. If we generalize equation (\ref{eq:bgt}) to
\be
g(s,\mu)=F{\mu^3\Omega^2\over \rho^2}w(s),
\ee
where $w'(s)$ is of order unity, then this condition becomes $\lambda\lesssim
2\pi F^{1/2} h$, where $h=\mu/\rho$ is the effective thickness, and we have
replaced the wavenumber $k$ by $2\pi/\lambda$. For typical values of $F$
(eq. \ref{eq:fdef}), the condition becomes $\lambda\lesssim 20h$. We expect
that our two-dimensional approximation remains valid so long as $\lambda\gtrsim
h$, so there is roughly one decade in wavenumber in which the instability
should be present whenever $w'(s_K)<0$. 

We have established that a strong short-wavelength instability is present in
rings in which $\p g/\p s < 0$, that is, in which the shear stress becomes
larger in absolute value when the shear rate becomes smaller in absolute
value. We now analyze the nonlinear behavior of this instability in a
ring model that is simplified even further. 

\section{A toy model}

\label{sec:toy}

In this section we investigate a simple fluid system that illustrates an
instability similar to the one we derived in the preceding section. Consider a
homogeneous two-dimensional system of viscous incompressible fluid with
surface density $\mu$. The system is infinite in the $y$-direction, and all of
its physical properties are assumed to be independent of $y$. The system is in
shear flow in the $y$-direction, described by the velocity field ${\bf
v}=(0,v(x,t))$. The system has periodic shearing boundary conditions in the
$x$-direction; that is, $v(x+\Delta x,t)=v(x,t)+s_K\Delta x$, where $\Delta x$
is the period in $x$, $s_K$ is the mean shear (i.e. the shear averaged over
length scales much larger than $\Delta x$). The only non-trivial component of
the Euler equation reads
\be
{\p v\over \p t}={1\over\mu}{\p\sigma_{xy}\over \p x},
\label{eq:eul}
\ee
where $\bf\sigma$ is the stress tensor. 

We let $s(x,t)\equiv\p v/\p x$ denote the shear. We shall assume that the
stress is a nonlinear function of the shear,
$\sigma_{xy}=g(s)$. Differentiating equation (\ref{eq:eul}) with respect to
$x$, we obtain
\be
{\p s\over \p t}={1\over\mu}{\p^2\over \p x^2}g(s).
\label{eq:ch}
\ee
The periodic boundary conditions require that
\be
\int_x^{x+\Delta x}s(x,t)dx=s_K\Delta x\qquad\hbox{for all $x$}.
\label{eq:bc}
\ee

Initially we assume that the system is in a state of uniform shear,
$s(x,t)=s_K$. A linearized analysis of equation (\ref{eq:ch}) reveals that
small disturbances of the form $\exp(\gamma t+ikx)$ satisfy the dispersion
relation
\be
\gamma =-{k^2g'(s_K)\over\mu},
\label{eq:instab}
\ee
exactly the same as the largest root in equation (\ref{eq:groot}). Thus we
have reproduced the instability derived in \S\ref{sec:linstab} in a simpler
system, in which we can more easily investigate the nonlinear evolution of the
instability.

The final state of an unstable system with $g'(s_K)<0$ is straightforward to
describe. First we define

\be
G(s)\equiv \int_0^sg(s')ds',\qquad F(t)\equiv \int_x^{x+\Delta x}G(s(x,t))dx.
\label{eq:gfdef}
\ee
Using equation (\ref{eq:ch}) and the boundary conditions, we have
\be
{dF\over dt}=\int_x^{x+\Delta x}g(s){\p s\over\p t}dx={1\over\mu}
\int_x^{x+\Delta x} g(s){\p^2\over\p x^2}g(s)\,dx=-{1\over\mu} 
\int_x^{x+\Delta x} \left(\p g\over\p x\right)^2dx \le 0.
\ee
Since $F$ decreases with time, the system must approach a steady state as
$t\to\infty$. The steady-state solutions of equation (\ref{eq:ch}) are those
in which $g(s)$ is a linear function of $x$; the periodic boundary conditions
require that the linear term vanishes, so $g(s)=\hbox{constant}$
in the steady
state. However, the only solution of this form that satisfies the mean shear
constraint (\ref{eq:bc}) is $s=s_K$, which is unstable. The resolution of this
apparent contradiction is that the final state of the unstable system is
piecewise constant in the shear $s$, alternating between states $s_-$ and
$s_+$ that satisfy $g(s_-)=g(s_+)=g(s_K)$. The fractions $f_\pm$ of the
spatial interval $\Delta x$ that are occupied by each of these two phases are
constrained by the relations
\be
f_-+f_+=1, \qquad f_-s_-+f_+s_+=s_K,
\label{eq:fracs}
\ee
where the second equality follows from equation (\ref{eq:bc}). However, the
equations provide no information on the spatial distribution of the two phases
in the final state; for example, the characteristic domain size is
unknown. 

The evolution from a uniform but unstable initial state into two distinct
phases is known as spinodal decomposition in studies of alloys and binary
fluids; in this case the shear $s$ is the concentration of one of the two
components, $g(s)$ is the chemical potential, the functional $F$ is the free
energy of the system, and equation (\ref{eq:bc}) corresponds to conservation
of mass \citep{bray94}. This conservation law places our model in the class of
phase transitions with a conserved order parameter.

The function $g(s)$ must satisfy several constraints in any physically
plausible system: (i) $g(s)$ is an odd function of $s$ (since the directions
$+y$ and $-y$ are equivalent); (ii) $g(0)=0$ (since there is no stress if
there is no shear); (iii) $g(s)$ has the same sign as $s$ (since the viscosity
is positive). A functional form that is general enough for our purposes is
\be
g(s)=s(as^2+b|s|+c);
\label{eq:gdef}
\ee
condition (iii) then requires that $a>0$, $c>0$, and $b>-(4ac)^{1/2}$.

\begin{figure}
%\epsscale{0.5}
\plottwo{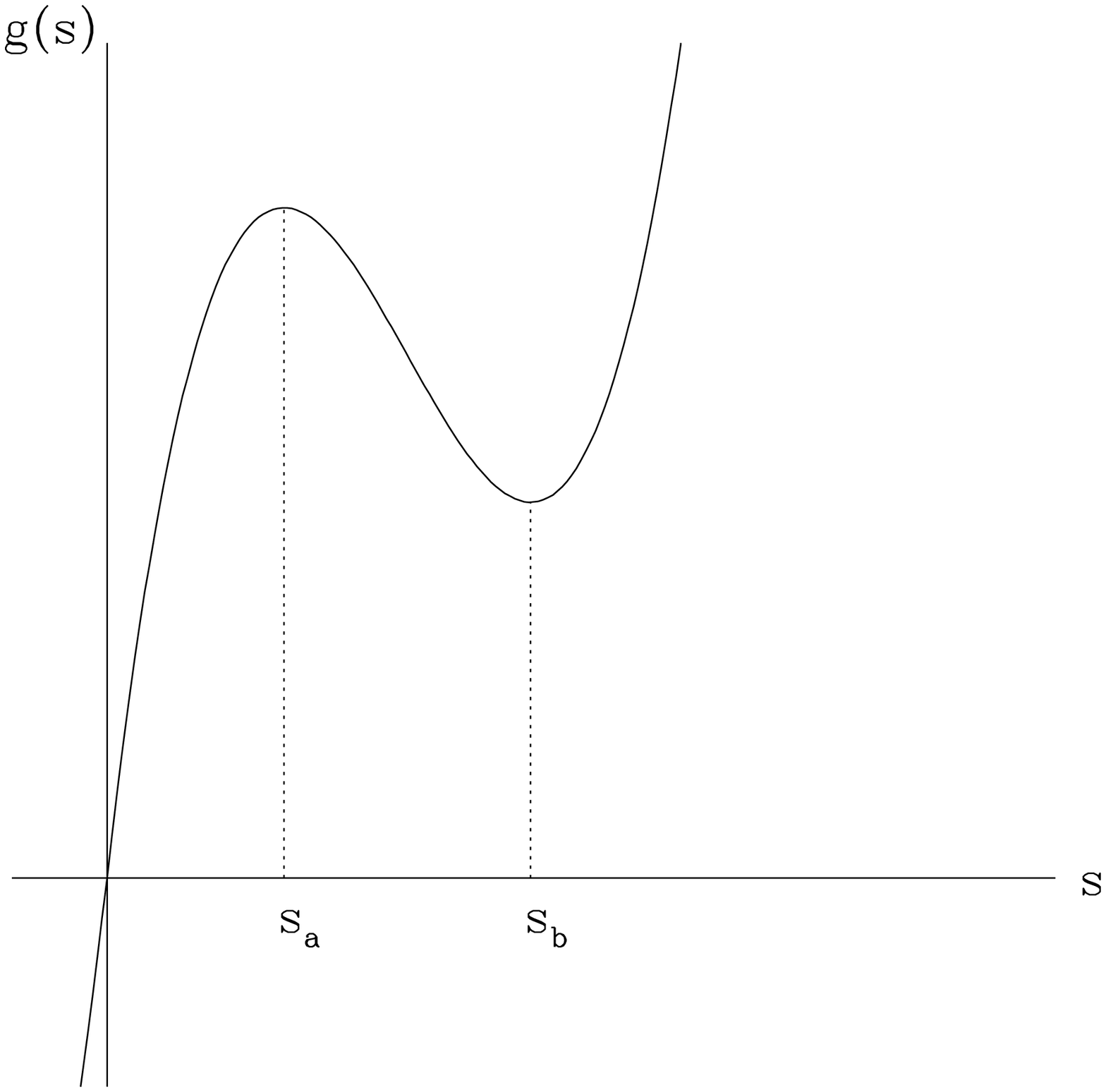}{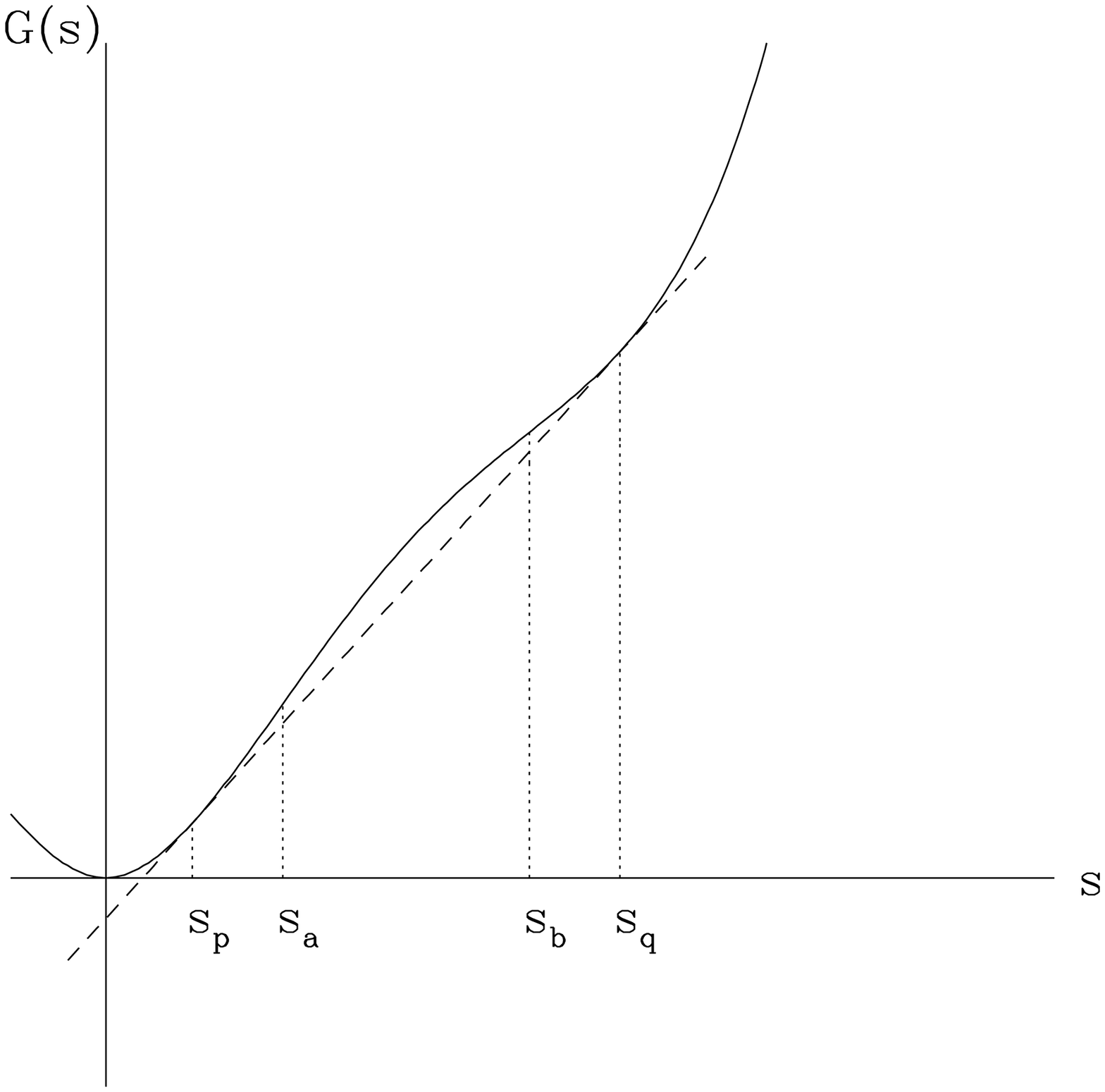}
\caption{(left panel) A possible form for the shear stress $g(s)$ as a
function of shear $s$, in the model of \S\ref{sec:toy}. Uniform shear is
unstable in the interval $s_a<s<s_b$ (eq.\ \ref{eq:instab}). (right panel) The
function $G(s)$ defined in equation (\ref{eq:gfdef}), shown for the same
parameters and on the same scale as the left panel. States in the interval
$s_p<s<s_a$ and $s_b<s<s_q$ are metastable. The values $s_p$ and $s_q$ are
determined by solving the equality corresponding to the inequality
(\ref{eq:meta}), along with the condition $g(s_p)=g(s_q)$. The solution of
these two simultaneous equations is equivalent to finding the straight line
tangent to $G(s)$ at two points.}
\label{fig:ggg}
\end{figure}

With this form for $g(s)$, there is an unstable region ($g'(s)<0$) if and only
if $b<-(3ac)^{1/2}$. When this condition is satisfied, all values of $|s|$
between $s_a>0$ and $s_b>0$ are unstable, where
\be
s_{a,b}=-{b\over 3a}\pm {(b^2-3ac)^{1/2}\over 3a};
\ee
The interval $s_a<|s|<s_b$ in which a uniform shear is unstable is called the
spinodal interval (Figure \ref{fig:ggg}).

There is also interesting behavior outside the spinodal interval. Suppose that
the system is in uniform shear $s_1$ outside the spinodal interval,
$s_1>s_b$. Suppose that there is another shear state outside the spinodal
interval, $s_2<s_a$, such that $g(s_1)=g(s_2)$. Then if a small element
$\delta x$ changes its shear state from $s_1$ to $s_2$, and the remaining
fluid increases its shear rate from $s_1$ to $s_1+\delta s_1$ so as to satisfy
the mean shear constraint (\ref{eq:bc}), the resulting change in free energy
is
\be
\delta F=\delta x[G(s_2)-G(s_1)-G'(s_1)(s_2-s_1)],
\ee
which is negative if
\be
G'(s_1)=g(s_1)> {G(s_2)-G(s_1)\over s_2-s_1}.
\label{eq:meta}
\ee
Shear states in which this inequality is satisfied---the interval $s_p<s<s_a$
and $s_b<s<s_q$ in Figure \ref{fig:ggg}---are metastable, because the mixture
of two distinct phases has a lower free energy.

Equation (\ref{eq:ch}) is ill-posed, since the growth rate of the instability
(\ref{eq:instab}) becomes extremely large for short-wavelength
disturbances. It is convenient, and physically plausible, to mitigate
this violent instability by modifying equation (\ref{eq:ch}) to
\be
{\p s\over \p t}={1\over\mu}{\p^2\over \p x^2}g(s)-\epsilon{\p^4 s\over\p
x^4},
\label{eq:cha}
\ee where $\epsilon$ is a small positive constant related to the thickness of
the interface. The additional term can be thought of as penalizing the growth
of solutions with large gradients.

We now investigate the properties of solutions of equation (\ref{eq:cha}).

It is simple to see that the mean shear is conserved,
\be
{d\over dt}\int_x^{x+\Delta x}s(x,t)dx=0,
\label{eq:cons}
\ee
which implies that if the mean shear condition (\ref{eq:bc}) is satisfied
initially then it is automatically satisfied for all time.

The dispersion relation for small disturbances is
\be
\gamma=-{k^2g'(s_K)\over\mu}-\epsilon k^4;
\label{eq:most}
\ee
thus only wavenumbers that satisfy $k^2<-g'(s_K)/(\epsilon\mu)$ are unstable;
the most unstable wavenumber is given by $k_{\rm max}^2=-\half
g'(s_K)/(\epsilon\mu)$.

Let us define
\be
 F(t)\equiv \int_x^{x+\Delta x}\left[{G(s(x,t))\over\mu}
 +\half\epsilon\left(\p s\over \p x\right)^2\right]dx,
\ee
where $G(s)$ is defined in equation (\ref{eq:gfdef}). It is straightforward to
show for periodic boundary conditions that
\be
{dF\over dt}=-\int_x^{x+\Delta x} \left({1\over\mu}{\p g\over\p
x}-\epsilon{\p^3s\over\p x^3}\right)^2dx \le 0.
\ee
Since $F$ decreases with time, the system must approach a steady state as
$t\to\infty$.

The steady-state solutions of equation (\ref{eq:cha}) are those in which
$g(s)/\mu-\epsilon(d^2 s/d x^2)$ is a linear function of $x$; the periodic
boundary conditions require that the linear term vanishes, so the steady state
solutions satisfy
\be
{d^2s\over dx^2}-{1\over\epsilon\mu}g(s)+p=0,
\ee
where $p$ is a constant. This equation is equivalent to motion in the
potential $-G(s)/(\epsilon\mu)+ps$, so there is an ``energy'' integral
\be
\half\left(ds\over dx\right)^2-{1\over\epsilon\mu}G(s)+ps=q;
\label{eq:energy}
\ee
the properties of the steady-state solutions can thus be determined by
examining the contours of the left-hand side of (\ref{eq:energy}) in the phase
plane with coordinates $(s,ds/dx)$.

We now introduce dimensionless variables,
\be
\xi\equiv \left(c\over\epsilon\mu\right)^{1/2}x, \qquad \tau\equiv
{c^2\over\epsilon\mu^2}t, \qquad y\equiv \left(a\over c\right)^{1/2}s;
\ee recall that $a$ and $c$ are positive. Equation (\ref{eq:cha}), with the
stress-shear relation (\ref{eq:gdef}), becomes
\be
{\p y\over \p \tau}={\p^2\over \p \xi^2}(y + By|y| + y^3)-{\p^4 y\over\p
\xi^4}, \qquad B\equiv {b\over (ca)^{1/2}}.
\label{eq:chb}
\ee The boundary conditions are periodic, with period $\Delta\xi\equiv
(c/\epsilon\mu)^{1/2}\Delta x$. Equation (\ref{eq:chb}) is the Cahn-Hilliard
equation, which has been widely used to model pattern formation in phase
transitions.

Consider small disturbances to uniform shear flow with shear $s_K$. The
dispersion relation for disturbances of the form $\exp(\gamma\tau+i\kappa\xi)$
is
\be
\gamma=-\kappa^2(1+2B|y_K|+3y_K^2) - \kappa^4,
\ee
where $y_K\equiv (a/c)^{1/2}s_K$. There is instability if $\gamma>0$ for some
$\kappa$, which occurs if
\be
B<-{1+3y_K^2\over 2|y_K|}.
\ee
If this instability is present, the stress as a function of shear $g(s)$ is
triple-valued: there are three roots $s_-<s_K<s_+$ to the equation
$g(s)=g(s_K)$. The root $s_K$ has $g'(s_K)<0$ and hence is unstable, while
$g'(s_\pm)>0$ so the roots $s_\pm$ are stable.

Almost all of the steady-state solutions of equation (\ref{eq:chb}) are
bounded and periodic, and can be expressed analytically in terms of Jacobian
elliptic functions \citet{ncs84}. However, these are {\em not} the final state
of the system: it turns out that all of the periodic solutions are unstable
\citep{car84}. The only stable, stationary solution is the ``kink'' solution,
\be
y(x)=\pm\ffrac{1}{3}B\pm\left(\ffrac{1}{3}B^2-1\right)^{1/2}
\tanh\left[\left(\ffrac{1}{3}B^2-1\right)^{1/2}{x\over\sqrt{2}}\right],
\label{eq:kink}
\ee
where the two $\pm$ signs are independent and we require that
$B<-3/\sqrt{2}$. Of course, this solution does not satisfy our periodic
boundary conditions.

Numerical integration of the partial differential equation (\ref{eq:chb}) for
a unstable initial state shows that the system evolves to a state in which the
shear is almost always nearly equal to either $s_-$ or $s_+$, just like the
solutions of the simpler equation (\ref{eq:ch}). Moreover, the interfaces
between high- and low-shear domains gradually drift, so that high-shear and
low-shear domains eventually coalesce. Once the distance between interfaces is
large compared to unity, the shape of the shear curve in the transition region
is given approximately by the kink solution (\ref{eq:kink}). Because of the
conservation law (\ref{eq:cons}), the interfaces cannot move independently.
Their interactions can be modeled by treating each interface as a particle
that exerts forces on other particles \citep{kaw82,mh95}. This process is
known as coarsening or ripening. In an infinite system, the coarsening process
continues indefinitely, although at a slower and slower rate as the domain
sizes grow. Thus complete thermodynamic equilibrium is never achieved.

The coarsening process in the one-dimensional Cahn-Hilliard equation exhibits
scaling behavior, that is, the characteristic distance between the interfaces
grows indefinitely but the correlation function retains the same
shape. However, in a planetary ring the coarsening will be halted when the
distance between interfaces becomes comparable to the tidal width of
\S\ref{sec:tensile}.  Thus our toy model leads naturally to the conclusion
that the irregular structure in Saturn's rings should have a characteristic
size comparable to the tidal width, which we estimated could be as large as
$\sim 100\km$.

Assuming that the characteristic width of the interface between solid and
liquid phases is $L$, the parameter $\epsilon$ is of order $L^2\nu$ where
$\nu$ is the kinematic viscosity. It is plausible that $L$ is comparable to
the radius of large ring particles.

One interesting unresolved issue is whether our one-dimensional approximation
is adequate. Coarsening in the Cahn-Hilliard equation is much faster for
systems with more than one dimension, where the free energy associated with
the interfaces or domain walls can be reduced by reducing their radius of
curvature. Is the modest curvature of annuli in the B-ring sufficient to make
this an important contributor to coarsening dynamics? A second unresolved
issue is the importance of noise. The Cahn-Hilliard equation is a
zero-temperature model; adding thermal noise converts the equation to a
finite-temperature model. Is such a model more relevant for Saturn's B ring,
and if so, what sets the effective temperature (impacts of small bodies?
gravitational wakes from large ring particles?)? Are the fluctuations
sufficient to trigger phase separation in the metastable region of the
Cahn-Hilliard equation?

\section{Discussion}

We have suggested that the irregular structure in Saturn's B ring arises from
the formation of solid ring-particle assemblies, which are limited in size
by the competition between tidal forces and the tensile strength of these
assemblies. We have shown that if the shear stress is a decreasing function of
the shear, $\p\Sigma_{xy}/\p s < 0$, then the ring is unstable at short
wavelengths. Using a toy model of an incompressible, non-Newtonian fluid in
shear flow, we have argued that this instability leads to the formation of
domains in which the shear takes on one of two values; it is natural to
identify one of these with the solid phase, and one with a liquid phase in
which the (absolute value of the) shear rate is greater than Keplerian. The
toy model is ill-posed at short wavelengths; if this is remedied by adding a
term that can be associated with the energy of domain walls (the term
proportional to $\epsilon$ in eq.\ \ref{eq:cha}), then the domains gradually
coarsen, leading to structure on scales much larger than the original unstable
wavelengths.

There are many shortcomings in this work. We have not derived a constitutive
relation for a ring-particle fluid with cohesion, and thus we have not shown
that the instability condition $\p\Sigma_{xy}/\p s < 0$ is satisfied in rings
with cohesive forces (although the simple kinetic-theory model of
\citet{bgt85} is marginally unstable; see eq.\ \ref{eq:bgt}). We have not
derived the statistical properties of the equilibrium ring structure, in which
the growth of larger domains due to coarsening is balanced by their
destruction from tidal forces. We have not discussed why the rigidly rotating
solid regions should have different phase functions or albedo from the regions
in which the particles are in differential rotation. We have not shown that
solid annuli are dynamically stable. Finally, we have not established that
ring particles stick to one another with the required cohesive strength;
laboratory experiments suggest that a frost layer is needed but we do not know
whether such layers are present in the B ring.

The most powerful tool we have to investigate at least some of these issues is
$N$-body simulations of local ring dynamics. Such simulations have already
provided considerable insight into the dynamics of rings composed of
inelastically colliding particles \citep{wt88,ric94,mos96,sss01}; what is
required is to generalize them to include cohesion.

\acknowledgments

I thank David Huse and Roman Rafikov for helpful discussions, and Ignacio
Mosqueira for detailed and thoughtful comments. This research was supported in
part by NASA grant NAG5-10456.

\end{document}